\documentclass[10pt,a4paper,onecolumn]{article}
\usepackage{marginnote}
\usepackage{graphicx}
\usepackage{xcolor}
\usepackage{authblk,etoolbox}
\usepackage{titlesec}
\usepackage{calc}
\usepackage{tikz}
\usepackage{hyperref}
\hypersetup{colorlinks,breaklinks,
            urlcolor=[rgb]{0.0, 0.5, 1.0},
            linkcolor=[rgb]{0.0, 0.5, 1.0}}
\usepackage{caption}
\usepackage{tcolorbox}
\usepackage{amssymb,amsmath}
\usepackage{ifxetex,ifluatex}
\usepackage{seqsplit}
\usepackage{fixltx2e} 
\usepackage[
  backend=biber,
]{biblatex}
\bibliography{paper.bib}

\usepackage[top=3.5cm, bottom=3cm, right=1.5cm, left=1.0cm,
            headheight=2.2cm, reversemp, includemp, marginparwidth=4.5cm]{geometry}

\titleformat{\section}
  {\normalfont\sffamily\Large\bfseries}
  {}{0pt}{}
\titleformat{\subsection}
  {\normalfont\sffamily\large\bfseries}
  {}{0pt}{}
\titleformat{\subsubsection}
  {\normalfont\sffamily\bfseries}
  {}{0pt}{}
\titleformat*{\paragraph}
  {\sffamily\normalsize}

\usepackage{fancyhdr}
\makeatletter
\let\ps@plain\ps@fancy
\fancyheadoffset[L]{4.5cm}
\fancyfootoffset[L]{4.5cm}

\definecolor{linky}{rgb}{0.0, 0.5, 1.0}

\newtcolorbox{repobox}
   {colback=red, colframe=red!75!black,
     boxrule=0.5pt, arc=2pt, left=6pt, right=6pt, top=3pt, bottom=3pt}

\newcommand{\ExternalLink}{%
   \tikz[x=1.2ex, y=1.2ex, baseline=-0.05ex]{%
       \begin{scope}[x=1ex, y=1ex]
           \clip (-0.1,-0.1)
               --++ (-0, 1.2)
               --++ (0.6, 0)
               --++ (0, -0.6)
               --++ (0.6, 0)
               --++ (0, -1);
           \path[draw,
               line width = 0.5,
               rounded corners=0.5]
               (0,0) rectangle (1,1);
       \end{scope}
       \path[draw, line width = 0.5] (0.5, 0.5)
           -- (1, 1);
       \path[draw, line width = 0.5] (0.6, 1)
           -- (1, 1) -- (1, 0.6);
       }
   }

\patchcmd{\@maketitle}{center}{flushleft}{}{}
\patchcmd{\@maketitle}{center}{flushleft}{}{}
\patchcmd{\@maketitle}{\LARGE}{\LARGE\sffamily}{}{}
\def\maketitle{{%
  
  \AB@maketitle}}
\makeatletter
\renewcommand\AB@affilsepx{ \protect\Affilfont}
\renewcommand\AB@affilnote[1]{{\bfseries #1}\hspace{3pt}}
\makeatother

\renewcommand\Affilfont{\sffamily\small\mdseries}
\ifnum 0\ifxetex 1\fi\ifluatex 1\fi=0 
  \usepackage[T1]{fontenc}
  \usepackage[utf8]{inputenc}

\else 
  \ifxetex
    \usepackage{mathspec}
  \else
    \usepackage{fontspec}
  \fi
  \defaultfontfeatures{Ligatures=TeX,Scale=MatchLowercase}

\fi
\IfFileExists{upquote.sty}{\usepackage{upquote}}{}
\IfFileExists{microtype.sty}{%
\usepackage{microtype}
\UseMicrotypeSet[protrusion]{basicmath} 
}{}

\usepackage{hyperref}
\hypersetup{unicode=true,
            pdftitle={pynucastro: an interface to nuclear reaction rates and code generator for reaction network equations},
            pdfborder={0 0 0},
            breaklinks=true}
\usepackage{graphicx,grffile}
\makeatletter
\def\maxwidth{\ifdim\Gin@nat@width>\linewidth\linewidth\else\Gin@nat@width\fi}
\def\maxheight{\ifdim\Gin@nat@height>\textheight\textheight\else\Gin@nat@height\fi}
\makeatother
\setkeys{Gin}{width=\maxwidth,height=\maxheight,keepaspectratio}
\IfFileExists{parskip.sty}{%
\usepackage{parskip}
}{
\setlength{\parindent}{0pt}
\setlength{\parskip}{6pt plus 2pt minus 1pt}
}
\ifx\paragraph\undefined\else
\let\oldparagraph\paragraph
\renewcommand{\paragraph}[1]{\oldparagraph{#1}\mbox{}}
\fi
\ifx\subparagraph\undefined\else
\let\oldsubparagraph\subparagraph
\renewcommand{\subparagraph}[1]{\oldsubparagraph{#1}\mbox{}}
\fi

\title{pynucastro: an interface to nuclear reaction rates and code generator
for reaction network equations}

        \author[1]{Donald E. Willcox}
          \author[1]{Michael Zingale}
    
      \affil[1]{Department of Physics and Astronomy, Stony Brook University}
  \date{\vspace{-5ex}}

\begin{document}
\maketitle

\marginpar{
  \sffamily\small

  {\bfseries DOI:} \href{https://doi.org/10.21105/joss.00588}{\color{linky}{10.21105/joss.00588}}

  \vspace{2mm}

  {\bfseries Software}
  \begin{itemize}
    \setlength\itemsep{0em}
    \item \href{https://github.com/openjournals/joss-reviews/issues/588}{\color{linky}{Review}} \ExternalLink
    \item \href{https://github.com/pynucastro/pynucastro}{\color{linky}{Repository}} \ExternalLink
    \item \href{http://dx.doi.org/10.5281/zenodo.1202434}{\color{linky}{Archive}} \ExternalLink
  \end{itemize}

  \vspace{2mm}

  {\bfseries Submitted:} 15 February 2018\\
  {\bfseries Published:} 18 March 2018

  \vspace{2mm}
  {\bfseries Licence}\\
  Authors of papers retain copyright and release the work under a Creative Commons Attribution 4.0 International License (\href{http://creativecommons.org/licenses/by/4.0/}{\color{linky}{CC-BY}}).
}

\section{Summary}\label{summary}

pynucastro addresses two needs in the field of nuclear astrophysics:
visual exploration of nuclear reaction rates or networks and automated
code generation for integrating reaction network ODEs. pynucastro
accomplishes this by interfacing with nuclear reaction rate
parameterizations published by the JINA Reaclib project (Cyburt et al.
2010).

Interactive exploration is enabled by a set of classes that provide
methods to visualize the temperature dependency of a rate, evaluate it
at a particular temperature, and find the exponent, n, for a simple
\(\rm{T^n}\) parameterization. From a collection of rates, the flow
between the nuclei can be visualized interactively using Jupyter
widgets. These features help both with designing a network for a
simulation as well as for teaching nuclear astrophysics in the
classroom.

After selecting a set of rates for a given problem, pynucastro can
construct a reaction network from those rates consisting of Python code
to calculate the ODE right hand side. Generated Python right hand sides
evolve species in the reaction network, and pynucastro includes a Python
example integrating the CNO cycle for hydrogen burning.

pynucastro can also generate Fortran code implementing reaction
networks, using SymPy (Meurer et al. 2017) to determine the system of
ODEs comprising the network. From the symbolic expressions for the ODE
right hand side, pynucastro also generates a routine to compute the
analytic Jacobian matrix for implicit integration.

Fortran networks incorporate weak, intermediate, and strong reaction
rate screening for the Reaclib rates (Graboske et al. 1973; Alastuey and
Jancovici 1978; Itoh et al. 1979). These networks can also include
selected weak reaction rate tabulations (Suzuki, Toki, and Nomoto 2016).
To calculate energy generation in Fortran networks, pynucastro uses
nuclear binding energies from the Atomic Mass Data Center (Huang et al.
2017; Wang et al. 2017) and the 2014 CODATA recommended values for the
fundamental physical constants (Mohr, Newell, and Taylor 2016).

pynucastro is capable of generating two kinds of Fortran reaction
networks. The first type is a standalone network with a driver program
to integrate species and energy generation using the variable-order ODE
integration package VODE (Brown, Byrne, and Hindmarsh 1989). This
Fortran driver program is designed to be easy to use and can integrate
reaction networks significantly faster than is possible for the
generated Python networks.

Secondly, pynucastro can generate a Fortran network consisting of right
hand side and Jacobian modules that evolve species, temperature, and
energy generation for the StarKiller Microphysics code. Via StarKiller
Microphysics, astrophysical simulation codes such as Castro (Almgren et
al. 2010) and Maestro (Nonaka et al. 2010) can directly use pynucastro
reaction networks. pynucastro includes a carbon burning network with
tabulated \(\rm{A=23}\) Urca weak reactions currently used for studying
white dwarf convection with Maestro (Zingale et al. 2017).

Future work will focus on implementing nuclear partition functions to
compute reverse reaction rates in the Reaclib library (Rauscher and
Thielemann 2000; Rauscher 2003). It is also in some cases necessary to
compute reverse reaction rates using detailed balance with a consistent
nuclear mass model instead of using the parameterized reverse reaction
rates in Reaclib (Lippuner and Roberts 2017). Additionally, work is
ongoing to port the networks generated for StarKiller Microphysics to
CUDA Fortran to support parallel reaction network integration on GPU
systems (Zingale et al. 2017). We intend to implement this port directly
into the pynucastro-generated networks.

We wish to thank Abigail Bishop for discussions about code generation
for the StarKiller Microphysics code as well as for exploratory
calculations. We are grateful to Max P. Katz for numerous discussions
that enabled the ongoing port of pynucastro-generated networks to CUDA
Fortran. We also wish to thank Christopher Malone for discussions about
various implementation details in pynucastro as well as sample code to
improve element identification. We especially thank Josiah Schwab for
helpful discussions about nuclear partition functions and reverse rates
as well as for testing pynucastro and pointing out issues in
visualization and documentation. This work was supported by DOE/Office
of Nuclear Physics grant DE-FG02-87ER40317.

\section*{References}\label{references}
\addcontentsline{toc}{section}{References}

\hypertarget{refs}{}
\hypertarget{ref-Screening.Alastuey.1978}{}
Alastuey, A., and B. Jancovici. 1978. ``Nuclear reaction rate
enhancement in dense stellar matter.'' \emph{The Astrophysical Journal}
226 (December): 1034--40.
doi:\href{https://doi.org/10.1086/156681}{10.1086/156681}.

\hypertarget{ref-castro}{}
Almgren, A. S., V. E. Beckner, J. B. Bell, M. S. Day, L. H. Howell, C.
C. Joggerst, M. J. Lijewski, A. Nonaka, M. Singer, and M. Zingale. 2010.
``CASTRO: A New Compressible Astrophysical Solver. I. Hydrodynamics and
Self-gravity.'' \emph{The Astrophysical Journal} 715 (June): 1221--38.
doi:\href{https://doi.org/10.1088/0004-637X/715/2/1221}{10.1088/0004-637X/715/2/1221}.

\hypertarget{ref-VODE.1989}{}
Brown, Peter N., George D. Byrne, and Alan C. Hindmarsh. 1989. ``VODE: A
Variable-Coefficient ODE Solver.'' \emph{SIAM Journal on Scientific and
Statistical Computing} 10 (5): 1038--51.
doi:\href{https://doi.org/10.1137/0910062}{10.1137/0910062}.

\hypertarget{ref-Reaclib.2010}{}
Cyburt, Richard H., A. Matthew Amthor, Ryan Ferguson, Zach Meisel, Karl
Smith, Scott Warren, Alexander Heger, et al. 2010. ``The JINA REACLIB
Database: Its Recent Updates and Impact on Type-I X-ray Bursts.''
\emph{The Astrophysical Journal Supplement Series} 189 (1): 240.
\url{https://doi.org/10.1088/0067-0049/189/1/240}.

\hypertarget{ref-Screening.Graboske.1973}{}
Graboske, H. C., H. E. Dewitt, A. S. Grossman, and M. S. Cooper. 1973.
``Screening Factors for Nuclear Reactions. II. Intermediate Screening
and Astrophysical Applications.'' \emph{The Astrophysical Journal} 181
(April): 457--74.
doi:\href{https://doi.org/10.1086/152062}{10.1086/152062}.

\hypertarget{ref-AME2016.1}{}
Huang, W.J., G. Audi, Meng Wang, F.G. Kondev, S. Naimi, and Xing Xu.
2017. ``The AME2016 atomic mass evaluation (I). Evaluation of input
data; and adjustment procedures.'' \emph{Chinese Physics C} 41 (3):
030002. \url{https://doi.org/10.1088/1674-1137/41/3/030002}.

\hypertarget{ref-Screening.Itoh.1979}{}
Itoh, N., H. Totsuji, S. Ichimaru, and H. E. Dewitt. 1979. ``Enhancement
of thermonuclear reaction rate due to strong screening. II - Ionic
mixtures.'' \emph{The Astrophysical Journal} 234 (December): 1079--84.
doi:\href{https://doi.org/10.1086/157590}{10.1086/157590}.

\hypertarget{ref-SkyNet.2017}{}
Lippuner, J., and L. F. Roberts. 2017. ``SkyNet: A Modular Nuclear
Reaction Network Library.'' \emph{The Astrophysical Journal Supplement
Series} 233 (December): 18.
doi:\href{https://doi.org/10.3847/1538-4365/aa94cb}{10.3847/1538-4365/aa94cb}.

\hypertarget{ref-SymPy.2017}{}
Meurer, Aaron, Christopher P. Smith, Mateusz Paprocki, Ondřej Čertík,
Sergey B. Kirpichev, Matthew Rocklin, Amit Kumar, et al. 2017. ``SymPy:
Symbolic Computing in Python.'' \emph{PeerJ Computer Science} 3
(January): e103.
doi:\href{https://doi.org/10.7717/peerj-cs.103}{10.7717/peerj-cs.103}.

\hypertarget{ref-CODATA.2014}{}
Mohr, Peter J., David B. Newell, and Barry N. Taylor. 2016. ``CODATA
recommended values of the fundamental physical constants: 2014.''
\emph{Rev. Mod. Phys.} 88 (3). American Physical Society: 035009.
doi:\href{https://doi.org/10.1103/RevModPhys.88.035009}{10.1103/RevModPhys.88.035009}.

\hypertarget{ref-MAESTRO:Multilevel}{}
Nonaka, A., A. S. Almgren, J. B. Bell, M. J. Lijewski, C. M. Malone, and
M. Zingale. 2010. ``MAESTRO:An Adaptive Low Mach Number Hydrodynamics
Algorithm for Stellar Flows.'' \emph{The Astrophysical Journal
Supplement Series} 188: 358--83.
\url{https://doi.org/10.1088/0067-0049/188/2/358}.

\hypertarget{ref-Rauscher2003}{}
Rauscher, T. 2003. ``Nuclear Partition Functions at Temperatures
Exceeding 10\(^{10}\) K.'' \emph{The Astrophysical Journal Supplement
Series} 147 (August): 403--8.
doi:\href{https://doi.org/10.1086/375733}{10.1086/375733}.

\hypertarget{ref-RaTh2000}{}
Rauscher, T., and F.-K. Thielemann. 2000. ``Astrophysical Reaction Rates
from Statistical Model Calculations.'' \emph{Atomic Data and Nuclear
Data Tables} 75 (May): 1--351.
doi:\href{https://doi.org/10.1006/adnd.2000.0834}{10.1006/adnd.2000.0834}.

\hypertarget{ref-Suzuki.2016}{}
Suzuki, Toshio, Hiroshi Toki, and Ken'ichi Nomoto. 2016.
``Electron-capture and \(\rm{\beta}\)-decay Rates for sd-Shell Nuclei in
Stellar Environments Relevant to High-density O--Ne--Mg Cores.''
\emph{The Astrophysical Journal} 817 (2): 163.
\url{https://doi.org/10.3847/0004-637x/817/2/163}.

\hypertarget{ref-AME2016.2}{}
Wang, Meng, G. Audi, F.G. Kondev, W.J. Huang, S. Naimi, and Xing Xu.
2017. ``The AME2016 atomic mass evaluation (II). Tables, graphs and
references.'' \emph{Chinese Physics C} 41 (3): 030003.
\url{https://doi.org/10.1088/1674-1137/41/3/030003}.

\hypertarget{ref-Zingale.astronum.2017}{}
Zingale, M., A. S. Almgren, M. G. Barrios Sazo, V. E. Beckner, J. B.
Bell, B. Friesen, A. M. Jacobs, et al. 2017. ``Meeting the Challenges of
Modeling Astrophysical Thermonuclear Explosions: Castro, Maestro, and
the AMReX Astrophysics Suite.'' \emph{ArXiv E-Prints}, November.
\url{http://adsabs.harvard.edu/abs/2017arXiv171106203Z}.

\end{document}